\pdfoutput=1  
\def\kx{{{\hat{S}}_x}}
\def\ky{{{\hat{S}}_y}}
\def\kz{{{\hat{S}}_z}}
\def\Lx{{{\hat{L}}_x}}
\def\Ly{{{\hat{L}}_y}}
\def\Lz{{{\hat{L}}_z}}

\def\nk{n_{\rm b}}

\def\dert#1#2{\frac{{{d}}{#1}}{{{d}}{#2}}}

\def\eqi{\begin{equation}}
\def\eqf{\end{equation}}
\def\eqia{\begin{eqnarray}}
\def\eqfa{\end{eqnarray}}
\def\Om{\mathit{\Omega}}
\def\rp#1#2{{#1\over#2}}
\def\lb#1{\label{#1}}
\def\kap{\bds{\hat{S}}}

\def\bds#1{\boldsymbol{#1}}

\def\cO{\cos\Om}
\def\sO{\sin\Om}
\def\cOO{\cos 2\Om}
\def\sOO{\sin 2\Om}

\def\cII{\cos 2I}

\def\ton#1{\left(#1\right)}
\def\qua#1{\left[#1\right]}
\def\grf#1{\left\{#1\right\}}

\documentclass[useAMS,usenatbib]{mn2e}
\usepackage{amsmath,textgreek,w-greek}
\usepackage{amscd}
\usepackage{amssymb}
\usepackage{graphicx,epsfig}
\usepackage{txfonts}
\usepackage{xr-hyper}

\RequirePackage{color}

\allowdisplaybreaks[1]

\title[WAPS-33 b parameters from orbital dynamics]{Accurate characterization of the stellar and orbital parameters of the exoplanetary system WASP-33 b from orbital dynamics}

\author[L. Iorio]{L.
Iorio$^{1}$\thanks{E-mail:
lorenzo.iorio@libero.it}\\
$^{1}$I Ministero dell'Istruzione, dell'Universit\`{a} e della
Ricerca (M.I.U.R.), Viale Unit\`{a} di Italia 68
Bari, (BA) 70125,
Italy}

\begin{document}

\maketitle

\label{firstpage}

\begin{abstract}
By using the most recently published Doppler tomography measurements and accurate theoretical modeling of the oblateness-driven orbital precessions, we tightly constrain some of the physical and orbital parameters of the planetary system hosted by the fast rotating star WASP-33. In particular, the measurements of the orbital inclination $i_{\rm p}$ to the plane of the sky and of the sky-projected spin-orbit misalignment $\lambda$ at two epochs \textcolor{black}{about} six years apart allowed for the determination of the longitude of the ascending node $\Omega$ and of the orbital inclination $I$ to the apparent equatorial plane at the same epochs. As a consequence, average rates of change $\dot\Omega_{\rm exp},~\dot I_{\rm exp}$ of this two orbital elements, accurate to a $\approx 10^{-2}~\textrm{deg}~\textrm{yr}^{-1}$  level, were calculated as well. By comparing them to general theoretical expressions  $\dot\Omega_{J_2},~\dot I_{J_2}$ for their precessions induced by an oblate star whose symmetry axis is arbitrarily oriented, we were able to determine the angle $i^{\star}$ between  the line of sight the star's spin  ${\bds S}^{\star}$ and its first even zonal harmonic $J_2^{\star}$ obtaining $i^{\star} \lb{ris_i} = \textcolor{black}{142}^{+10}_{-11}~\textrm{deg},~J_2^{\star} \lb{ris_j2} = \ton{2.1^{+0.8}_{-0.5}}\times 10^{-4}.$ As a by-product, the angle between ${\bds S}^{\star}$ and the orbital angular momentum $\bds L$ is as large as about $\psi \approx 100$ deg $\ton{\psi^{2008} = 99^{+5}_{-4}~\textrm{deg},~\psi^{\textcolor{black}{2014}} = 103^{+5}_{-4}~\textrm{deg}}$, and changes at a rate $\dot\psi = 0.\textcolor{black}{7}^{+1.5}_{-1.6}~\textrm{deg}~\textrm{yr}^{-1}$. The predicted general relativistic Lense-Thirring precessions, or the order of $\approx 10^{-3}~\textrm{deg}~\textrm{yr}^{-1}$, are, at present, about one order of magnitude below the measurability threshold.
\end{abstract}


%

\section{Introduction}
Steady observations of a test particle orbiting its primary over time intervals much longer than its orbital period $P_{\rm b}$ can reveal peculiar cumulative features of its orbital motion which may turn out to be  valuable tools to either put to the test fundamental theories or characterize the physical properties of the central body acting as source of the gravitational field. It has been just the case so far in several different astronomical and astrophysical scenarios ranging, e.g., from the early pioneering determinations of the multipole moments of the non-central gravitational potential of the Earth with artificial satellites \citep{1961AJ.....66....8K, 1962GeoJ....6..270K, 1962RPPh...25...63C} to the celebrated corroborations of the Einsteinian General Theory of Relativity (GTR) with the explanation of the anomalous (at that time) perihelion precession of Mercury \citep{1915SPAW...47..831E}-observationally known since decades \citep{LeVer859}-, several binary systems hosting at least one emitting pulsar \citep{1975ApJ...195L..51H, 2003Natur.426..531B, 2004Sci...303.1153L, 2006Sci...314...97K}, and  Earth's satellites  \citep{2010PhRvL.105w1103L, 2014PhRvD..89h2002L} implementing earlier ideas put forth since the dawn of the space era and beyond \citep{1954PASP...66...13L, 1978A&A....69..321C}. Plans exist to use in a similar way the stars revolving around the supermassive black hole in Sgr A$^{\ast}$ \citep{2008ApJ...689.1044G, 2009ApJ...692.1075G, 2010ApJ...720.1303A, 2015ApJ...809..127Z}.

With over\footnote{See, e.g., http://exoplanets.org/ on the    WEB.} 1500 planets discovered so far and counting  \citep{2014PASP..126..827H}, most of which orbiting very close to their parent stars \citep{2013Sci...340..572H}, extrasolar systems \citep{2014exha.book.....P}, in principle, represent ideal probes to determine or, at least, constrain some physical parameters of their stellar partners through their orbital dynamics. One of them is the quadrupole mass moment $J_2$, accounting for the flattening of the star. It  is connected with fundamental properties of the stellar interior such as, e.g., the non-uniform distribution for both velocity rates and mass \citep{2009ApJ...703.1791R, 2011JASTP..73..241D, 2011EPJH...36..407R, 2013SoPh..287..161R}.
Also GTR may turn out a valuable goal for exoplanets' analysts also from a practical point of view. Indeed, by assuming its validity, it may be used as a tool for dynamically characterizing the angular momentum $\bds S$ of the host stars via the so-called Lense-Thirring effect \citep{1918PhyZ...19..156L}. Such a dynamical variable is able to provide relevant information about the inner properties of stars and their activity. Furthermore, it plays the role of an important diagnostic for putting to the test theories of stellar formation. The angular momentum can also have a crucial impact in stellar evolution, in particular towards the higher mass \citep{1971Ap&SS..13..234T, 1982ApJ...252..322W, 1990SoPh..128..287V, 1997PASP..109..759W, 2005ApJ...633..967H, 2005ApJS..156..245J}.
As a naive measure of the relevance of the Einsteinian theory of gravitation in a given binary system characterized by mass $M$, proper angular momentum $\bds S$ and extension $r$, the magnitude  of the ratios  of some typical gravitational lengths to $r$ can be assumed. By taking \citep{2003ASSL..293.....B}
\begin{align}
r_M \lb{rM} & = \rp{GM}{c^2}, \\ \nonumber \\
r_S \lb{rS} & = \rp{S}{M c},
\end{align}
where $G$ and $c$ are the Newtonian gravitational constant and the speed of light in vacuum, respectively, it can be easily noted that, for exoplanets hosted by Sun-like stars at, say, $r=0.005~\textrm{au}$, Eqs \ref{rM} to \ref{rS} yield
\begin{align}
\rp{r_M}{r} & = 2\times 10^{-6},\\ \nonumber \\
\rp{r_S}{r} & = 4\times 10^{-7}.
\end{align}
Such figures are substantially at the same  level of, or even larger than those of the double pulsar \citep{2003Natur.426..531B, 2004Sci...303.1153L, 2006Sci...314...97K}, for which one has
\begin{align}
\rp{r_M}{r} & = 4\times 10^{-6},\\ \nonumber \\
\rp{r_S}{r} & = 8\times 10^{-8}.
\end{align}
 It shows that, in principle, some of the extrasolar planetary systems may well represent important candidates to perform also tests of relativistic orbital dynamics.

In the present work, we will deal with WASP-33 b \citep{2010MNRAS.407..507C}. It is a planet closely transiting a fast rotating and oblate main sequence star along a circular, short-period ($P_{\rm b} = 1.21~\textrm{d}$) orbit which is highly inclined to the stellar equator. In \citet{2011Ap&SS.331..485I} it was suggested that, in view of the relatively large size of some classical and general relativistic orbital effects, they could be used to better characterize its parent star as long as  sufficient accurate data records were available. It has, now, became possible in view of the latest Doppler tomography measurements processed by \citet{2015ApJ...810L..23J}, and of more accurate theoretical models of the orbital precessions involved \citep{2011PhRvD..84l4001I, 2012GReGr..44..719I}.

The plan of the paper is as follows. In Section \ref{modello}, we illustrate our general analytical expressions for the averaged classical and relativistic precessions of some Keplerian orbital elements in the case of an arbitrary orientation of the stellar symmetry axis and of an unrestricted orbital geometry. Section \ref{coor} describes the coordinate system adopted in this astronomical laboratory.
 Our theoretical predictions of the orbital rates of change are compared to the corresponding phenomenologically measured precessions in Section \ref{legami}, where tight constraints on some key stellar parameters are inferred, and the perspectives of measuring the Lense-Thirring effect are discussed. Section \ref{finale} is devoted to summarizing our findings.
\section{The mathematical model of the orbital precessions}\lb{modello}
A particle at distance $r$ from a central rotating body of symmetry axis direction $\kap=\grf{\kx,\ky,\kz}$ experiences an additional non-central acceleration \citep{2005CeMDA..91..217V}
\eqi{\bds A}_{J_2} = -\rp{3GMJ_2R^2}{2r^4}\grf{\qua{1 - 5 \ton{\bds{\hat{r}}\bds\cdot\kap}^2}\bds{\hat{r}} + 2\ton{\bds{\hat{r}}\bds\cdot\kap} \kap  }\lb{AJ2},\eqf
which causes long-term orbital precessions. For a generic orientation of $\kap$ in a given coordinate system, they were analytically worked out by\footnote{The replacement $Q_2\rightarrow GM J_2 R^2$ must be done in the equations by \citet{2011PhRvD..84l4001I} to obtain the present ones. Other conventions exist in the literature about dimensional quadrupole moments $Q$, mainly differing for the sign and the inclusion of $G$.} \citet{2011PhRvD..84l4001I}. Among them\footnote{Also the argument of pericenter $\omega$ and the mean anomaly $\mathcal{M}$ are impacted by $J_2$ with long-term precessions. We will not display them here because they are not relevant in the present study.}, we have
\begin{align}
\dot\Omega_{J_2}  \lb{dOdtJ2} \nonumber & = \rp{3\nk J_2 R^2}{4 a^2\ton{1-e^2}^2}\grf{2\kz\cII\csc I\ton{\kx\sO - \ky\cO} + \right. \\ \nonumber \\
& +\left. \cos I\qua{1 - 3 \kz^2 +\ton{\ky^2 - \kx^2}\cOO  - 2\kx\ky\sOO}  }, \\ \nonumber \\
\dot I_{J_2} \lb{dIdtJ2} \nonumber & = -\rp{3\nk J_2 R^2}{2 a^2\ton{1-e^2}^2}\ton{\kx\cO + \ky\sO}\qua{\kz\cos I + \right. \\ \nonumber \\
& + \left.\sin I\ton{\kx\sO - \ky\cO} },
\end{align}
which will be relevant for our purposes.
In Eqs \ref{dOdtJ2} to \ref{dIdtJ2}, $a$ is the semimajor axis, $\nk=\sqrt{GM a^{-3}}$ is the Keplerian mean motion, $e$ is the eccentricity, $I$ is the inclination of the orbital plane with respect to the coordinate $\grf{x,y}$ plane adopted, and $\Omega$ is the longitude of the ascending node counted in the $\grf{x,y}$ plane from a reference $x$ direction to the intersection of the orbital plane with the $\grf{x,y}$ plane itself.
Note that if the body's equatorial plane is assumed as $\grf{x,~y}$ plane, i.e. if $\kx=\ky=0,~\kz=1$, \textcolor{black}{Eqs \ref{dOdtJ2} to \ref{dIdtJ2} reduce to the well known expressions \citep{2003ASSL..293.....B}
\begin{align}
\dot\Omega^{(0)}_{J_2} \lb{stan} &= -\rp{3\nk J_2 R^2}{2 a^2\ton{1-e^2}^2}\cos I, \\ \nonumber\\
\dot I^{(0)}_{J_2}  &= 0;
\end{align}
}
with this particular choice, $I$ coincides with the angle $\psi$ between $\bds S$ and the particle's orbital angular momentum $\bds L$. It is important to stress that, in the general case, the cumbersome multiplicative geometrical factor in Equation \ref{dOdtJ2} depending on the spatial orientation of the orbit and of the spin axis  does not reduce to $\cos\psi$, as it will explicitly turn out clear in Section \ref{coor}. On the other hand, it can be easily guessed from the fact that $\cos\psi$ is linear in the components of $\kap$, while the acceleration of Equation \ref{AJ2} is quadratic in them, whatever parametrization is adopted. Such an extrapolation of a known result valid only in specific cases is rather widespread in the literature (see, e.g., \citet{2011Ap&SS.331..485I, 2013ApJ...774...53B, 2015ApJ...810L..23J}), and may lead to errors when accurate results are looked for.
Eqs \ref{dOdtJ2} to \ref{dIdtJ2} are completely general, and can be used with any coordinate system provided that the proper identifications pertaining the angular variables are made.

The general relativistic gravitomagnetic field due to the angular momentum $\bds S$ of the central body induces the Lense-Thirring effect \citep{1918PhyZ...19..156L}, whose relevant orbital precessions, valid for an arbitrary orientation of $\bds S$, are\footnote{The gravitomagnetic pericenter precession, not relevant for us here, will not be shown.} \citep{2012GReGr..44..719I}
\begin{align}
\dot\Omega_S \lb{dOdtLT} & = \rp{2GS}{c^2 a^3\ton{1-e^2}^{3/2}}\qua{\kz + \cot I\ton{\ky\cO - \kx\sO} }, \\ \nonumber \\
\dot I_S \lb{dIdtLT} & = \rp{2GS}{c^2 a^3\ton{1-e^2}^{3/2}}\ton{\kx\cO + \ky\sO}.
\end{align}
%
\textcolor{black}{In the special case in which $\bds S$ is directed along the reference $z$ axis, Eqs \ref{dOdtLT} to \ref{dIdtLT} reduce to the textbook results \citep{2013CEJPh..11..531R}
\begin{align}
\dot\Omega_S^{(0)} & = \rp{2GS}{c^2 a^3\ton{1 - e^2}^{3/2}},\\ \nonumber \\
\dot I^{(0)}_S & = 0.
\end{align}
}
The perspectives of detecting general relativity, mainly in its spin-independent, Schwarzschild-type manifestations, with exoplanets have been studies so far by several authors \citep{2006NewA...11..490I, 2006ApJ...649..992A, 2006ApJ...649.1004A, 2006IJMPD..15.2133A, 2007MNRAS.377.1511H, 2008MNRAS.389..191P, 2008ApJ...685..543J, 2009ApJ...698.1778R, 2011MNRAS.411..167I, 2011Ap&SS.331..485I, 2011Ap&SS.332..107H, 2014MNRAS.438.1832X, 2012Ap&SS.341..323L, 2013RAA....13.1231Z}.
\section{The coordinate system adopted}
\begin{figure*}
\centerline{
\vbox{
\begin{tabular}{cc}
\epsfysize= 7.0 cm\epsfbox{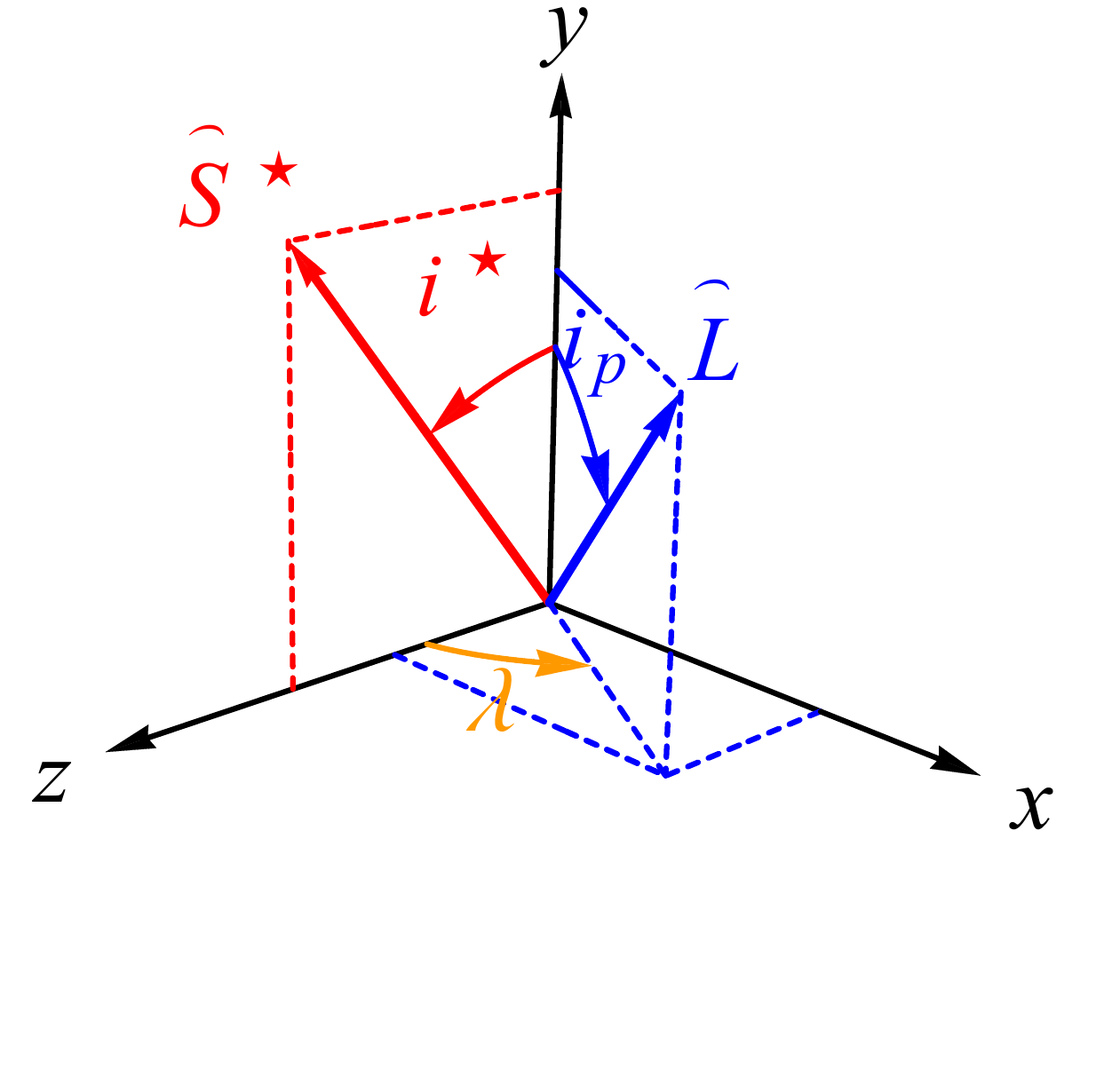} & \epsfysize= 7.0 cm\epsfbox{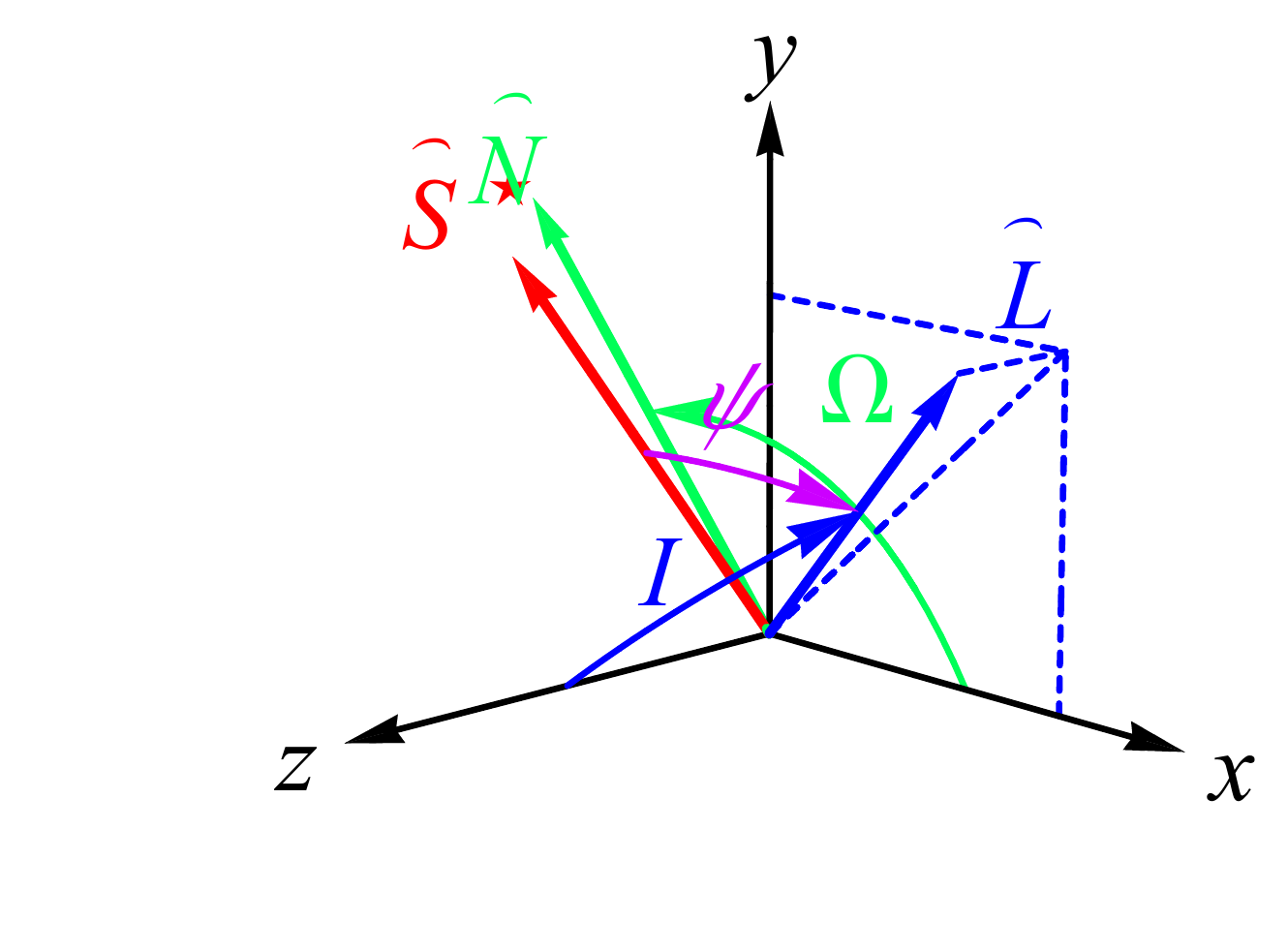}\\
\end{tabular}
}
}
\caption{The coordinate system adopted. The  axes $x$ and $z$  span the plane of the sky in such a way that the projection of the stellar spin  ${\bds S}^{\star}$ onto it defines the $z$ axis. The $y$ axis is directed along the line of sight towards the observer. The $\grf{x,y}$ plane is the apparent equatorial plane. The inclination of the orbital plane to the plane of the sky is $i_{\rm p}$, while $i^{\star}$ is the angle between the line of sight and the star's spin axis. The sky-projected spin-orbit misalignment angle $\lambda$ lies in the plane of the sky, and is delimited by the projections of ${\bds S}^{\star}$ and $\bds L$ onto it.  The unit vector $\bds{\hat{N}}$ of the line of the nodes lies in the apparent equatorial plane perpendicularly to the projection of $\bds L$ onto it. The longitude of the ascending node $\Omega$ is counted in the $\grf{x,~y}$ plane from the $x$ axis to the line of the nodes. The inclination of the orbital plane to the apparent equatorial plane is $I$. The angle between  ${\bds{\hat{S}}}^{\star}$ and $\bds{\hat{L}}$ is $\psi$. The values of the angles used to produce the picture were arbitrarily chosen just for illustrative purposes; they do not correspond to the actual configuration of WASP-33 b.}\label{figura1}
\end{figure*}

\lb{coor}
For consistency reasons with the conventions adopted by \citet{2015ApJ...810L..23J}, who, in turn, followed \citet{2000A&A...359L..13Q}, the coordinate system used in the present analysis is as follows (see Figure \ref{figura1}).
%
%
%

The line of sight, directed towards the observer, is assumed as reference $y$ axis, while the $z$ axis is determined by the projection of the stellar spin axis ${\kap}^{\star}$ onto the plane of the sky\textcolor{black}{, which is inferred from observations}. The $x$ axis is straightforwardly chosen perpendicular to both the other two axes in such a way to form a right-handed coordinate system; it generally does not point towards the Vernal Equinox $\curlyvee$ at a reference epoch. With the present choice, the coordinate $\{x,~y\}$ plane does not coincide with the plane of the sky which, instead, is now spanned by the $z$ and $x$ axes; the $\grf{x,~y}$ plane is known as apparent equatorial plane \citep{2000A&A...359L..13Q}. The planetary longitude of the ascending node $\Omega$ lies in it, being counted from the $x$ axis to the intersection of the orbital plane with the apparent equatorial plane itself; thus, in general, $\Omega$ does not stay in the plane of the sky.
Moreover, with  such conventions, the angle $I$  between the orbital plane and the coordinate $\{x,~y\}$ plane entering  Eqs \ref{dOdtJ2} to \ref{dIdtJ2} and Eqs \ref{dOdtLT} to \ref{dIdtLT} is not the orbital inclination $i_{\rm p}$, which refers the plane of the sky \textcolor{black}{and is one of the orbital parameters directly accessible to observations}. Instead, $I$, which is also the angle  \textcolor{black}{from the unit vector $\bds{k}$ of the $z$ axis to the planetary orbital angular momentum $\bds L$}, has to be identified with the angle $\alpha$ of \citet{2000A&A...359L..13Q}. By considering it as a colatitude angle of $\bds{\hat{L}}$ in a spherical coordinate system,
the components of the unit vector of the planetary orbital angular momentum are
\begin{align}
\Lx \lb{LxO} &=  \sin I\sO, \\ \nonumber \\
\Ly \lb{LyO}& = -\sin I\cO, \\ \nonumber \\
\Lz & \lb{LzO}= \cos I.
\end{align}
In view of Eqs \ref{LxO} to \ref{LzO},
\begin{align}
\dert{\bds{\hat{L}}}t & = \rp{\dot\Omega^{(0)}_{J_2}}{\cos I}\ton{\bds{\hat{L}}\bds\cdot\bds{\hat{S}}}\bds{\hat{S}}\bds\times\bds{\hat{L}}, \\ \nonumber \\
\dert{\bds{\hat{L}}}t & = \dot\Omega^{(0)}_{S}~\bds{\hat{S}}\bds\times\bds{\hat{L}},
\end{align}
concisely summarize Eqs \ref{dOdtJ2} to \ref{dIdtJ2} and Eqs \ref{dOdtLT} to \ref{dIdtLT}.
\textcolor{black}{Another} angle which is measurable is the projected spin-orbit misalignment $\lambda$. It lies in the plane of the sky, and is delimited by the projections of both the stellar spin axis and of the planetary orbital angular momentum. In our coordinate system, $\lambda,~i_{\rm p}$ are the longitude and the colatitude spherical angles, respectively, with $\lambda$ reckoned from the $z$ axis to the projection of $\bds{\hat{L}}$ onto the plane of the sky. As such, the components of the planetary orbital angular momentum versor can also be written as
\begin{align}
\Lx \lb{Lxl} & = \sin i_{\rm p}\sin\lambda, \\ \nonumber \\
\Ly \lb{Lyl}& = \cos i_{\rm p}, \\ \nonumber \\
\Lz \lb{Lzl} & = \sin i_{\rm p}\cos \lambda.
\end{align}

\textcolor{black}{In general, both $I$ and $\Omega$, which explicitly enter Eqs \ref{dOdtJ2} to \ref{dIdtJ2} and Eqs \ref{dOdtLT} to \ref{dIdtLT}, are not directly measurable; they must be expressed in terms of the observable angles $i_{\rm p},~\lambda$. To this aim, it is useful to use the unit vector $\bds{\hat{N}}$ directed along the line of the nodes towards the ascending node, which is defined as \eqi\bds{\hat{N}} = \rp{\bds k\bds\times\bds{\hat{L}}}{\left|\bds k\bds\times\bds{\hat{L}}\right|}.\lb{Nvec}\eqf From Eqs \ref{Lxl} to \ref{Lzl}, its components are
\begin{align}
{\hat{N}}_x \lb{Nxl} & = -\rp{\cos i_{\rm p}}{\sqrt{\cos^2 i_{\rm p} + \sin^2 i_{\rm p}\sin^2\lambda } }, \\ \nonumber \\
{\hat{N}}_y \lb{Nyl} & =\rp{\sin i_{\rm p}\sin\lambda}{\sqrt{\cos^2 i_{\rm p} + \sin^2 i_{\rm p}\sin^2\lambda } }, \\ \nonumber \\
{\hat{N}}_z \lb{Nzl} & = 0.
\end{align}
Eqs \ref{LxO} to \ref{LzO} and the definition of Equation \ref{Nvec} allow to express the components of $\bds{\hat{N}}$ in terms of $I,~\Omega$ as
\begin{align}
{\hat{N}}_x \lb{NxO} & = \cos\Omega, \\ \nonumber \\
{\hat{N}}_y \lb{NyO} & =\sin\Omega, \\ \nonumber \\
{\hat{N}}_z \lb{NzO} & = 0.
\end{align}
By adopting the convention\footnote{\textcolor{black}{Instead, it seems that \citet{2015ApJ...810L..23J} follow the convention $-\uppi\leq\Omega\leq\uppi$.}} $0\leq\Omega\leq 2\uppi$, Equation \ref{NxO} yields\footnote{\textcolor{black}{Recall that the function $\arccos $ returns values from 0 to $\uppi$.}}
\begin{align}
\Omega \lb{Omega1}&= \arccos {\hat{N}}_x~\textrm{for}~{\hat{N}}_y \geq 0, \\ \nonumber \\
\Omega \lb{Omega2}&= 2\uppi - \arccos {\hat{N}}_x~\textrm{for}~{\hat{N}}_y < 0,
\end{align}
where ${\hat{N}}_x,~{\hat{N}}_y$ are expressed in terms of $i_{\rm p},~\lambda$ by means of Eqs \ref{Nxl} to \ref{Nyl}.
The inclination $I$, defined in the range $0\leq I\leq \uppi$, is obtained in terms of $i_{\rm p},~\lambda$ from
\eqi I=\arccos\ton{\bds k\bds\cdot\bds{\hat{L}}}\lb{Inc}\eqf and Eqs \ref{Lxl} to \ref{Lzl}.
}
%
%
%
%

If $i^{\star}$ is the angle between  from the line of sight to ${\kap}^{\star}$, the components of the star's spin axis in our coordinate system are
\begin{align}
\kx \lb{kx} &= 0, \\ \nonumber \\
\ky & = \cos i^{\star}, \\ \nonumber \\
\kz \lb{kz} & = \sin i^{\star}.
\end{align}

The angle $\psi$ between the stellar angular momentum ${\bds S}^{\star}$ and the planetary orbital angular momentum $\bds L$ can be computed from Eqs \ref{kx} to \ref{Lzl} as
\eqi\kap\bds\cdot\bds{\hat{L}}=\cos\psi = \cos i_{\rm p}\cos i^{\star} + \sin i_{\rm p}\sin i^{\star}\cos \lambda,\lb{cospsi}\eqf
in agreement with, e.g., \citet{2009ApJ...696.1230F, 2011Ap&SS.331..485I}.
Incidentally, Equation \ref{cospsi}, along with an analogous one which could be straightforwardly obtained from Eqs \ref{LxO} to \ref{LzO} and Eqs \ref{kx} to \ref{kz} in terms of $I,~\Omega$, explicitly shows that the node precession cannot be generally proportional to $\cos\psi$, as previously remarked in Section \ref{modello}.

\textcolor{black}{
Finally, the configurations $\grf{i_{\rm p},~\lambda,~i^{\star}}$ and $\grf{\uppi - i_{\rm p},~-\lambda,~\uppi - i^{\star}}$ are physically equivalent since they correspond to looking at the planetary system from the opposite sides of the plane of the sky \citep{2015ApJ...805...28M}. In both case, the angle $\psi$ remains the same, as explicitly shown by Equation \ref{cospsi}. According to Eqs \ref{dOdtJ2} to \ref{dIdtJ2}, the node precession remains unaltered, while the rate of $I$ changes by the amount
\begin{align}
\Delta \dot I_{J_2} \lb{DI}\nonumber &\doteq \dot I_{J_2}^{\grf{i_{\rm p},~\lambda,~i^{\star}}}-  \dot I_{J_2}^{\grf{\uppi-i_{\rm p},~-\lambda,~\uppi-i^{\star}}}=\\ \nonumber \\
&=\pm\rp{3\nk J_2 R^2\left|\sin\lambda\right|\cos i^{\star}\sin i_{\rm p}\cos\psi}{a^2\ton{1-e^2}^2\sqrt{\cos^2 i_{\rm p} + \sin^2 i_{\rm p}\sin^2\lambda}}.
\end{align}
In calculating Equation \ref{DI}, we used both Equation \ref{Omega1} and Equation \ref{Omega2} in $\dot I_{J_2}^{\grf{i_{\rm p},~\lambda,~i^{\star}}},~\dot I_{J_2}^{\grf{\uppi-i_{\rm p},~-\lambda,~\uppi-i^{\star}}}$ since the transformation $\grf{i_{\rm p},~\lambda}\rightarrow\grf{\uppi - i_{\rm p},~-\lambda}$ changes the sign of ${\hat{N}}_y$, as shown by Equation \ref{Nyl}. The $+$ sign in Equation \ref{DI} corresponds to using Equation \ref{Omega2}  in $\dot I_{J_2}^{\grf{i_{\rm p},~\lambda,~i^{\star}}}$ and Equation \ref{Omega1} in $\dot I_{J_2}^{\grf{\uppi-i_{\rm p},~-\lambda,~\uppi-i^{\star}}}$ while the $-$ sign is for Equation \ref{Omega1}  in $\dot I_{J_2}^{\grf{i_{\rm p},~\lambda,~i^{\star}}}$ and Equation \ref{Omega2} in $\dot I_{J_2}^{\grf{\uppi-i_{\rm p},~-\lambda,~\uppi-i^{\star}}}$.
}
\section{Constraining the stellar spin axis and oblateness}\lb{legami}
\subsection{Using the precessions of $I$ and $\Omega$}
Generally speaking, while the magnitude of the classical precessions driven by the star's oblateness is at the $\approx~\textrm{deg}~\textrm{yr}^{-1}$ level, the relativistic gravitomagnetic ones about three orders of magnitude smaller. Despite this discrepancy, if, on the one hand, the current state-of-the-art in the orbital determination of WASP-33 b \citep{2015ApJ...810L..23J}, based on data records $\textcolor{black}{5.89}$ years long \textcolor{black}{(from Nov 12, 2008 to Oct 4, 2014)}, does not yet allow for a measurement of the relativistic effects, on the other hand, they might exceed the measurability threshold in a not so distant future. Indeed, they are just $\approx~4-8$ times smaller than the present-day errors, which amount to $\approx 2-8\times 10^{-2}$ deg yr$^{-1}$ \citep{2015ApJ...810L..23J} for the node.

In the following, we will reasonably assume that the measured orbital precessions of WASP-33 b are entirely due to the star's oblateness. This will alow us to put much tighter constraints on either $i^{\star}$ and $J_2^{\star}$. Our approach is as follows.

The lucky availability of the measurements of both $i_{\rm p}$  and $\lambda$ at two different epochs some years apart leads to the calculation of the unobservable orbital parameters $\Omega~,I$ from Eqs \ref{Omega1} to \ref{Inc} at the same epochs.
\textcolor{black}{According to the measured values of $i_{\rm p},~\lambda$ by \citet{2015ApJ...810L..23J}, it is ${\hat{N}}_y<0$, so that Equation \ref{Omega2} must be used yielding
\begin{align}
\Omega^{2008} \lb{O08} & = 266.4^{+0.5}_{-0.2}~\textrm{deg}, \\ \nonumber \\
\Omega^{2014} \lb{O14} & =  268.58^{+0.04}_{-0.03}~\textrm{deg}.
\end{align}
The values by  \citet{2015ApJ...810L..23J} differ from Eqs \ref{O08} to \ref{O14} by $\uppi$, likely due to the different convention adopted for the node.
Since Equation \ref{Inc} returns
\begin{align}
I^{2008} \lb{I08} & = 110.0^{+0.5}_{-0.4}~\textrm{deg}, \\ \nonumber \\
I^{2014} \lb{I14} & =  112.9^{+0.2}_{-0.7}~\textrm{deg},
\end{align}
Eqs \ref{LxO} to \ref{LzO} and Eqs \ref{Lxl} to \ref{Lzl} agree both in magnitude and in sign.
 }
It is straightforward to compute the average rates of change $\dot\Omega_{\rm exp},~\dot I_{\rm exp}$ by simply taking the ratios of the differences $\Delta\Omega,~\Delta I$ of their values at the measurement's epochs to the time span, which in our case is $\Delta t= \textcolor{black}{5.89}~\textrm{yr}$. Our results are in Table \ref{tavola1}.
\begin{table*}
\centering
\caption{Measured and derived parameters for the WASP-33 system according to Table 1 of \citet{2015ApJ...810L..23J} and the present study. Our values for $\Omega,~I$ were inferred by \textcolor{black}{assuming $0\leq\Omega\leq 2\uppi$}, and calculating \textcolor{black}{Eqs \ref{Omega2} to \ref{Inc}} with the measured values of the orbital inclination $i_{\rm p}$ and the sky-projected spin-orbit misalignment angle $\lambda$ released by \citet{2015ApJ...810L..23J}, while the errors were found by numerically determining the maxima and minima of Eqs \ref{Omega2} to \ref{Inc} thought as functions of $i_{\rm p},~\lambda$ varying in the rectangle delimited by their measurement errors as per Table 1 of \citet{2015ApJ...810L..23J}. The same procedure was adopted for the errors in $\dot\Omega_{\rm exp},~\dot I_{\rm exp}$, assumed as a function of $\Omega^{2008},~\Omega^{2014},~I^{2008},~I^{2014}$ varying in the rectangle determined by the errors in them previously calculated. The time span adopted for calculating the precessions is $\Delta t= 5.89~\textrm{yr}$. \textcolor{black}{The different values of the node quoted by \citet{2015ApJ...810L..23J} with respect to ours are likely due to a different convention adopted by them for the ascending node.}}
\label{tavola1}
\begin{tabular}{lllll}
\noalign{\smallskip}
\hline
Parameter  & \citep{2015ApJ...810L..23J}  & This study  \\
\hline
$i^{2008}_{\rm p}$ & $86.61^{+0.46}_{-0.17}$~deg &  \textcolor{black}{Same}\\ \\
$i^{2014}_{\rm p}$ & $88.695^{+0.031}_{-0.029}$~deg & \textcolor{black}{Same} \\ \\
$\lambda^{2008}$ & $-110.06^{+0.40}_{-0.47}$~deg & \textcolor{black}{Same}\\ \\
$\lambda^{2014}$ & $-112.93^{+0.23}_{-0.21}$~deg & \textcolor{black}{Same}\\ \\
$\Omega^{2008}$ & $86.39^{+0.49}_{-0.18}$~deg & $\textcolor{black}{266.4}^{+0.5}_{-0.2}$~deg \\ \\
$\Omega^{2014} $ & $88.584^{+0.034}_{-0.032}$~deg & $\textcolor{black}{268.58}^{+0.04}_{-0.03}$~deg \\ \\
$\dot\Omega_{\rm exp}$  & $0.373^{+0.031}_{-0.083}$~deg~yr$^{-1}$ & $0.3\textcolor{black}{7}^{+0.04}_{-0.09}$~deg~yr$^{-1}$\\ \\
$I^{2008}$ & -- & $110.0^{+0.5}_{-0.4}$~deg \\ \\
$I^{2014} $ & -- & $112.9^{+0.2}_{-0.7}$~deg \\ \\
$\dot I_{\rm exp}$  & -- & $0.5^{+0.1}_{-0.2}$~deg~yr$^{-1}$\\ \\
$J_2^{\star}$  & $\qua{0.54,~3.5}\times 10^{-2}$  & $\ton{2.1^{+0.8}_{-0.5}}\times 10^{-4}$\\ \\
$i^{\star}$  & $\qua{11.22,~168.77}$~deg & $\textcolor{black}{142}^{+10}_{-11}~\textrm{deg}$\\ \\
$\psi^{2008}$ & -- & $99^{+5}_{-4}$~deg \\ \\
$\psi^{2014} $ & -- & $103^{+5}_{-4}$~deg \\ \\
$\dot \psi_{\rm exp}$  & -- & $0.\textcolor{black}{7}^{+1.5}_{-1.6}$~deg~yr$^{-1}$\\ \\
\hline
\end{tabular}
\end{table*}

Eqs \ref{dOdtJ2} to \ref{dIdtJ2} provide us with  an accurate mathematical model of the oblateness-driven precessions which, in view of its generality, can be straightforward applied to the present case. Eqs \ref{dOdtJ2} to \ref{dIdtJ2} can be viewed as two functions of the two independent variables $i^{\star},~J_2^{\star}$. By allowing them to vary within their physically admissible ranges \citep{2011Ap&SS.331..485I}, it is possible to equate $\dot\Omega_{J_2^{\star}},~\dot I_{J_2^{\star}}$ to $\dot\Omega_{\rm exp},~\dot I_{\rm exp}$ by obtaining certain stripes in the $\grf{i^{\star},~J_2^{\star}}$ plane whose widths are fixed by the experimental ranges of the observationally determined precessions quoted in Table \ref{tavola1}. If our model is correct and if it describes adequately the empirical results, the two stripes must overlap somewhere in the considered portion of the  $\grf{i^{\star},~J_2^{\star}}$ plane by determining an allowed region of admissible values for the inclination of the stellar spin axis to the line of sight and the star's dimensionless quadrupole mass moment.
%
%
\begin{figure*}
\centerline{
\vbox{
\begin{tabular}{cc}
\epsfysize= 8.0 cm\epsfbox{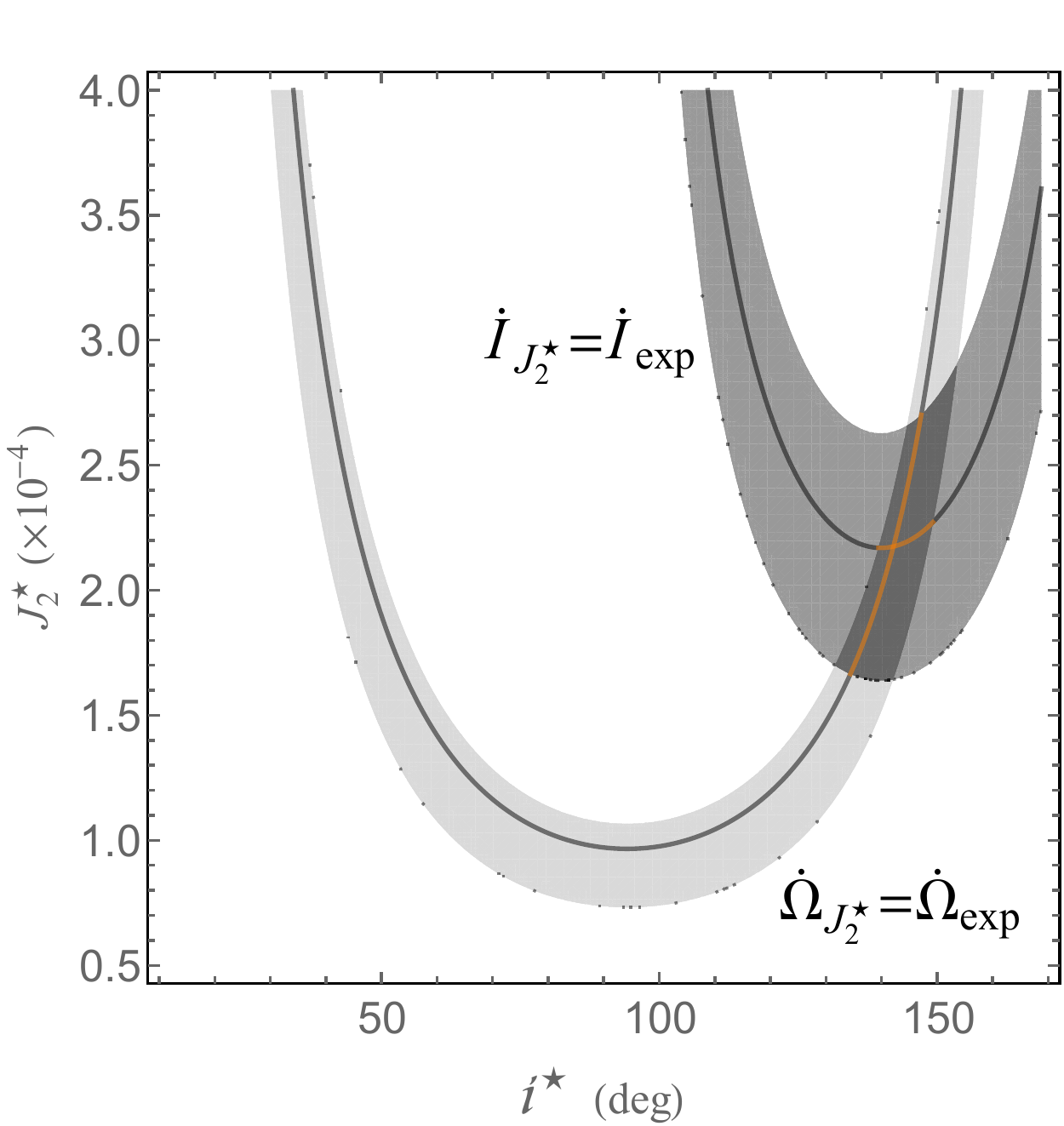} & \epsfysize= 8.0 cm\epsfbox{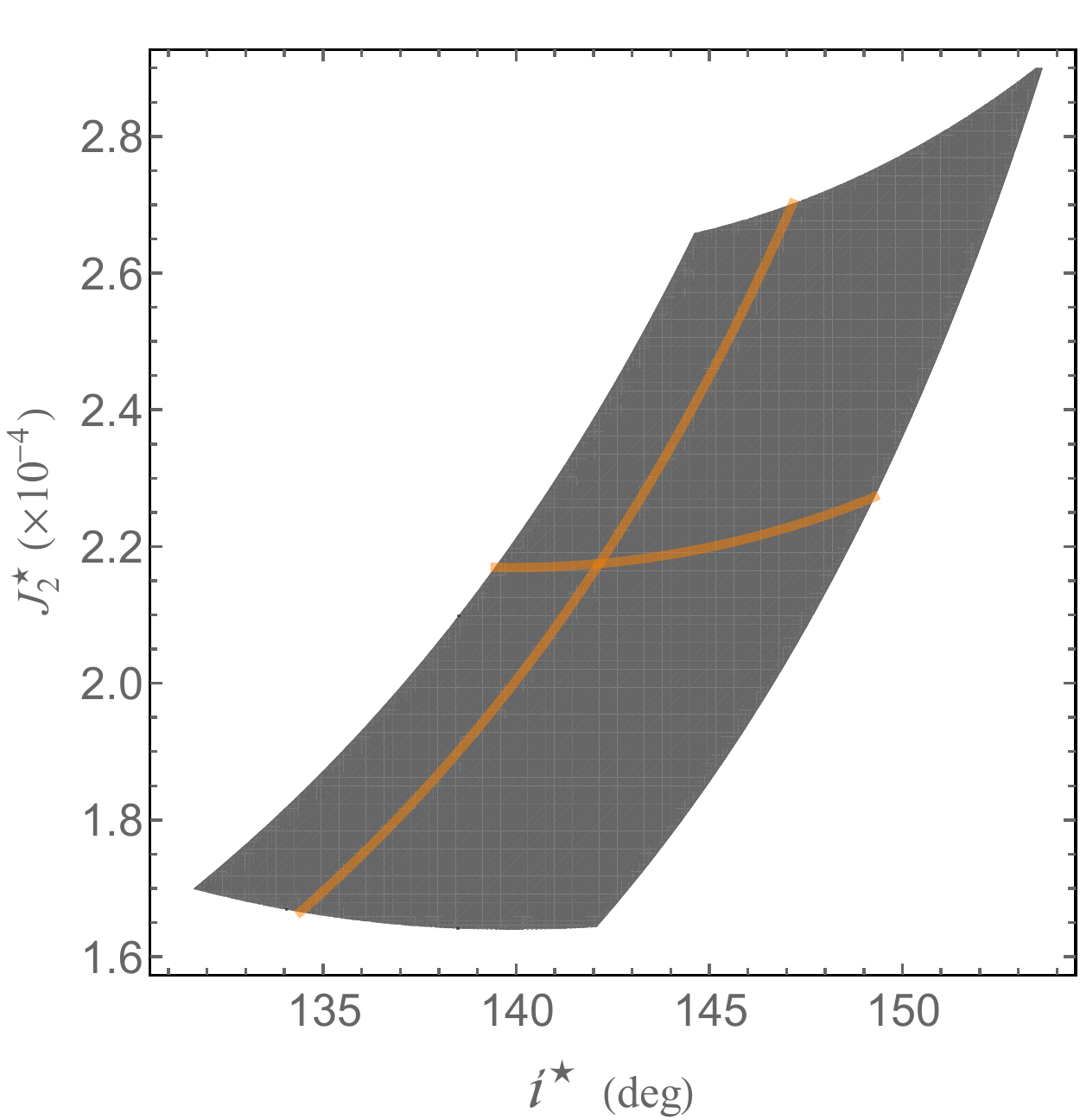}\\
\epsfysize= 8.0 cm\epsfbox{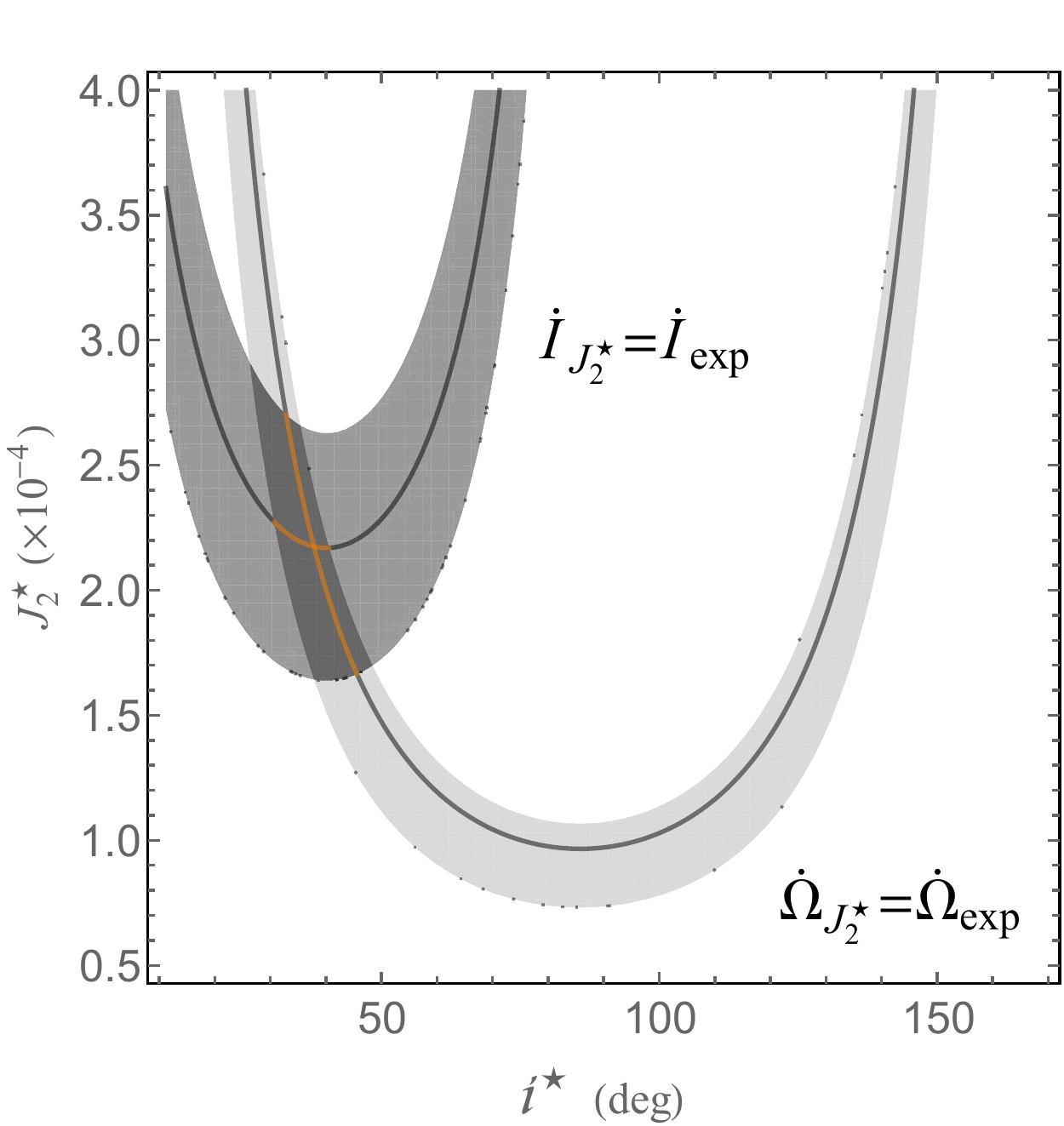} & \epsfysize= 8.0 cm\epsfbox{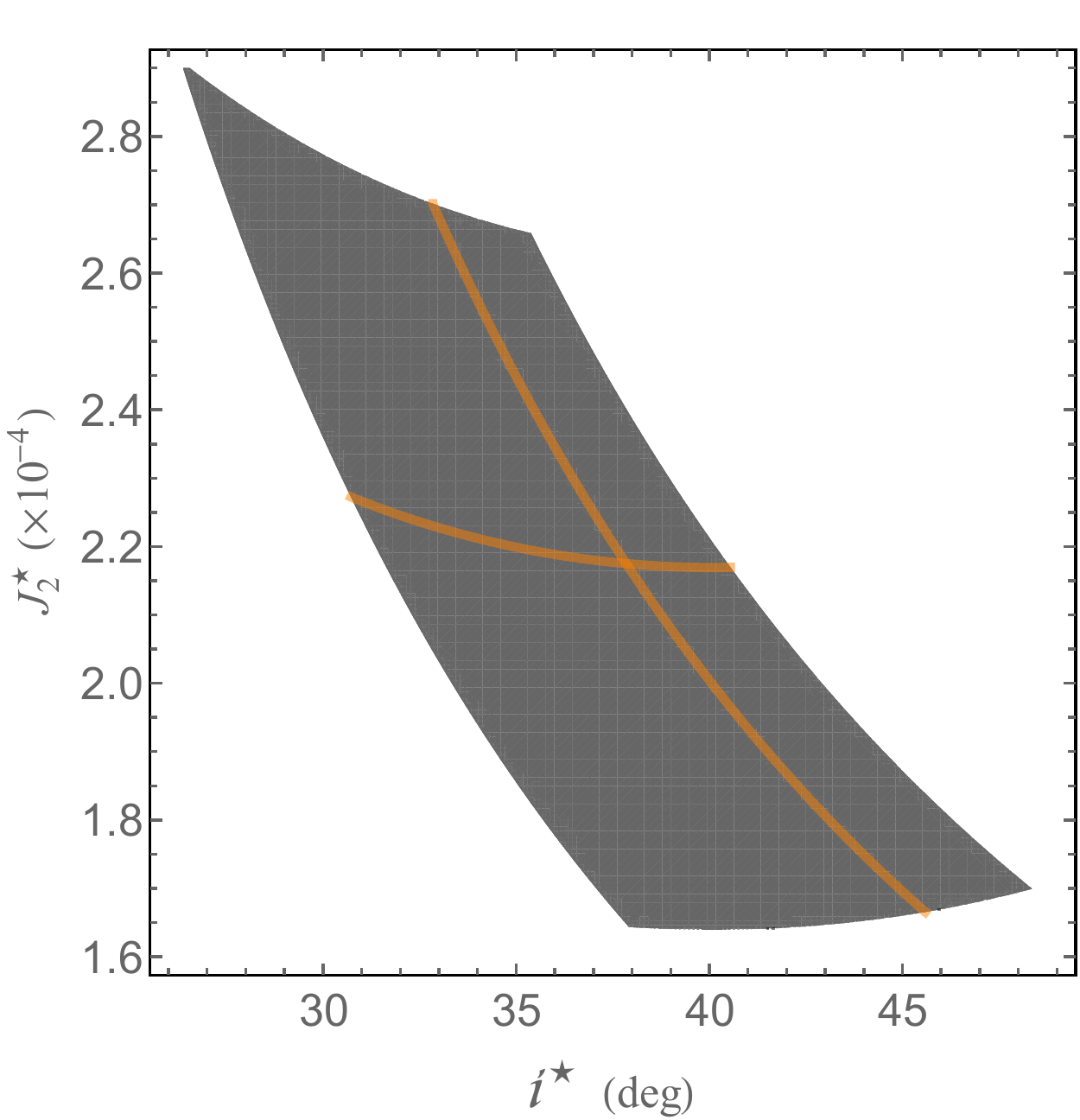}\\
\end{tabular}
}
}
\caption{\textcolor{black}{Upper row: }the darkest region in the plot is the experimentally allowed area in the $\grf{i^{\star},~J_2^{\star}}$ plane, which is enlarged in the right panel. It is determined by the overlapping of the permitted shaded stripes set by the precessions of the node $\Omega$ and the orbital inclination to the apparent equatorial plane $I$. We assumed that the experimental precessions $\dot \Omega_{\rm exp},~\dot I_{\rm exp}$  are entirely due to the stellar oblateness $J_2^{\star}$, within the experimental errors. For  $\dot \Omega_{J_2^{\star}},~\dot I_{J_2^{\star}}$, we used the mathematical model of Eqs \ref{dOdtJ2} to \ref{dIdtJ2} calculated with the values quoted in Table \ref{tavola1}; the values for $R^{\star},~M^{\star},~a$ were taken from \citet{2010MNRAS.407..507C}. The curves inside the shaded areas correspond to the best estimates for $\dot \Omega_{\rm exp},~\dot I_{\rm exp}$; their intersection   is given by $i^{\star} = \textcolor{black}{142}~\textrm{deg},~J_2^{\star} = 2.1\times 10^{-4}$. \textcolor{black}{Lower row: same as in the upper row, but with $\uppi-i_{\rm p},~-\lambda$. Note that the stripe for $\dot I$ is different, in agreement with Equation \ref{DI}. The solution for the stellar spin axis inclination corresponds to $\uppi-i^{\star}$. }}\label{figura2}
\end{figure*}
It is just the case, as depicted in \textcolor{black}{the upper row of} Figure \ref{figura2}. From it, it turns that
\begin{align}
i^{\star} \lb{ris_i}& = \textcolor{black}{142}^{+10}_{-11}~\textrm{deg},\\ \nonumber \\
J_2^{\star} \lb{ris_j2}& = \ton{2.1^{+0.8}_{-0.5}}\times 10^{-4}.
\end{align}
As a consequence, the angle between the orbital plane and the stellar equator and its precession is as reported in Table \ref{tavola1}.

\textcolor{black}{The lower row of Figure \ref{figura2} depicts the physically equivalent case with $\uppi-i_{\rm p},~-\lambda$. Now, ${\hat{N}}_y>0$, and Equation \ref{Omega1} must be used yielding
\begin{align}
\Omega^{2008} \lb{OO08} & = 86.4^{+0.5}_{-0.2}~\textrm{deg}, \\ \nonumber \\
\Omega^{2014} \lb{OO14} & =  88.58^{+0.04}_{-0.03}~\textrm{deg}.
\end{align}
While the stripe for $\dot\Omega$ is the same, it is not so for $\dot I$, as expected from Equation \ref{DI}; the intersection between the $\dot I,~\dot\Omega$ curves corresponds to \eqi\uppi - i^{\star} = {38}^{+10}_{-11}~\textrm{deg}.\eqf
It must be noted that $J_2^{\star}$ is unchanged.
 }
\subsection{Constraining the oblateness of Kepler-13 Ab}
An opportunity to apply the present method to another exoplanet is offered by Kepler-13 Ab, also known as KOI-13.01 \citep{2012MNRAS.421L.122S, 2014ApJ...788...92S, 2014ApJ...790...30J, 2015ApJ...805...28M}. By using the values of its physical\footnote{Contrary to WASP-33 b, in the case of Kepler-13 Ab also $i^{\star}$ is available \citep{2015ApJ...805...28M}.} and orbital parameters determined with the gravity darkened transit light curves and other observations \citep{2015ApJ...805...28M}, it is possible to compute analytically the rate of change of $\cos i_{\rm p}$ \textcolor{black}{in terms of $\dot\Omega,~\dot I$ by means of Equation \ref{LyO} and Equation \ref{Lyl},} and compare it to its accurately measured value \citep{2015ApJ...805...28M} in order to infer $J_2^{\star}$. We obtain
\eqi J_2^{\star} = \ton{6.0 \pm 0.6}\times 10^{-5}, \lb{risul}\eqf
in agreement with \citet{2015ApJ...805...28M} who seemingly used a different dynamical modelization. We calculated our uncertainty with a straightforward error propagation in our analytical expression of $J_2^{\star}$ thought as a function of the parameters $d|\cos i_{\rm p}|/dt,~\cos i_{\rm p},~i^{\star},~P_{\rm b},~a/R_{\star},~\lambda$ affected by experimental uncertainties \citep{2014ApJ...788...92S, 2015ApJ...805...28M}.

\textcolor{black}{The definition of the impact parameter
\eqi b = \ton{\rp{a}{R_{\star}}}\cos i_{\rm p},\lb{impa}\eqf
valid for a circular orbit, along with Equation \ref{LyO} and Equation \ref{Lyl},} allows us to use also the value of $\dot b_{\rm exp}$ independently measured by \citet{2012MNRAS.421L.122S} with the transit duration variation, although it is accurate only to $27\%$. We get
\eqi J_2^{\star} = \ton{8.6 \pm 2.4}\times 10^{-5}, \eqf
which is not in disagreement with Equation \ref{risul}.

From \textcolor{black}{Equation \ref{impa}}, it turns out that the analytical expressions of $\dot b$ and $d\cos i_{\rm p}/dt$ are not independent, so that the availability of independently measured values for both of them do not allow to determine/constrain any further dynamical effect with respect to $J_2^{\star}$. Luckily, it seems that other precessions, independent of $\dot b,~d\cos i_{\rm p}/dt$,  should be measurable via Doppler tomography in the next years or so \citep{2015ApJ...810L..23J, 2015ApJ...805...28M}. Depending on the final accuracy reached, such an important measurement will allow, at least in principle, to dynamically measure or, at least, constrain also the stellar spin by means of  the Lense-Thirring effect through\textcolor{black}{, e.g., $\dot\lambda$} calculated with Eqs \ref{dOdtLT} to \ref{dIdtLT}.
\section{Summary and conclusions}\lb{finale}
The use of a general model of the orbital precessions caused by the primary's oblateness,  applied to recent phenomenological measurements of some planetary orbital parameters of WASP-33 b taken at different epochs $\textcolor{black}{5.89}$ years apart, allowed us to tightly constrain the inclination $i^{\star}$ of the spin  ${\bds S}^{\star}$ of WASP-33 to the line of sight and its dimensionless quadrupole mass moment $J_2^{\star}$. Our analytical expressions are valid for arbitrary orbital geometries and spatial orientations of the body's symmetry axis.

By comparing our theoretical orbital rates of change of the longitude of the ascending node $\Omega$ and of the inclination $I$ of the orbital plane with respect to the apparent equatorial plane with the observationally determined ones, we obtained
$i^{\star} \lb{ris_i} = \textcolor{black}{142}^{+10}_{-11}~\textrm{deg},~J_2^{\star} \lb{ris_j2} = \ton{2.1^{+0.8}_{-0.5}}\times 10^{-4}.$ Furthermore, the angle between the stellar and orbital angular momenta at different epochs is $\psi^{2008} = 99^{+5}_{-4}~\textrm{deg},~\psi^{\textcolor{black}{2014}} = 103^{+5}_{-4}~\textrm{deg}$. Thus, it varies at a rate $\dot\psi = 0.\textcolor{black}{7}^{+1.5}_{-1.6}~\textrm{deg}~\textrm{yr}^{-1}$.

In view of the fact that WASP-33 b should transit its host star until 2062 or so and of the likely improvements in the measurement accuracy over the years, such an extrasolar planet will prove a very useful tool for an increasingly accurate  characterization  of the key physical and geometrical parameters of its parent star via its orbital dynamics. Moreover, also the determination of the general relativistic Lense-Thirring effect, whose predicted size is currently just one order of magnitude smaller than the present-day accuracy level in determining the planetary orbital precessions, may become a realistic target to be pursued over the next decades.

Furthermore, in view of its generality, our approach can be straightforwardly applied to any other exoplanetary system, already known or still to be discovered, for which at least the same parameters as of WASP-33 b are or will become accessible to the observation. A promising candidate, whose orbital precessions should be measurable via Doppler tomography in the next years, is Kepler-13 Ab. For the moment, we applied our method to it by exploiting its currently known parameters, and we were able to constrain its oblateness in agreemement with the bounds existing in the literature.

Finally, in principle, also the periastron, if phenomenologically measurable at different epochs as in the present case, can become a further mean to investigate the characteristics of highly eccentric exoplanetary systems-and to test general relativity as well-along the guidelines illustrated here.
\section*{Acknowledgements}
I would like to thank M.~C. Johnson for useful correspondence and clarifications. \textcolor{black}{I am indebted also to an anonymous referee for her/his efforts to improve the manuscript.}

\bibliography{exoplanetsbib}{}

\end{document}